1
2
# Genetic progression and the waiting time to cancer

Niko Beerenwinkel[1*], Tibor Antal[1], David Dingli[1], Arne Traulsen[1], Kenneth W. Kinzler[2], Victor E. Velculescu[2], Bert Vogelstein[2], Martin A. Nowak[1]

[1]Program for Evolutionary Dynamics, Harvard University, One Brattle Square, Cambridge, MA 02138, USA

[2]Ludwig Center and Howard Hughes Medical Institute, Sidney Kimmel Comprehensive Cancer Center at Johns Hopkins, Baltimore, MD 21231, USA

[*]Corresponding author:
phone: +1 (617) 496-5551, fax: +1 (617) 496-4629, email: beerenw@fas.harvard.edu

**Abstract**

**Background:** Cancer results from genetic alterations that disturb the normal cooperative behavior of cells. Recent high-throughput genomic studies of cancer cells have shown that the mutational landscape of cancer is complex and that individual cancers may evolve through mutations in as many as 20 different cancer-associated genes.

**Methodology and Principle Findings:** We use data published by Sjöblom et al. (2006) to develop a new mathematical model for the somatic evolution of colorectal cancers. We employ the Wright-Fisher process for exploring the basic parameters of this evolutionary process and derive an analytical approximation for the expected waiting time to the cancer phenotype. Our results highlight the relative importance of selection over both the size of the cell population at risk and the mutation rate. The model predicts that the observed genetic diversity of cancer genomes can arise under a normal mutation rate if the average selective advantage per mutation is on the order of 1%. Increased mutation rates due to genetic instability would allow even smaller selective advantages during tumorigenesis.

**Conclusions:** The complexity of cancer progression can be understood as the result of multiple sequential mutations, each of which has a relatively small but positive effect on net cell growth.

**Introduction**

The current view of cancer is that tumorigenesis is due to the accumulation of mutations in oncogenes, tumor suppressor genes, and genetic instability genes [1]. Sequential mutations in these genes, lead to most of the hallmarks of cancer [2]. Cancer research has benefited immensely from studies of uncommon inherited cancer syndromes that served to highlight the importance of individual genes in tumorigenesis [3]. Theoretical considerations have suggested that a handful of mutations, perhaps as few as three, may be sufficient for developing colorectal cancer [4,5]. This relatively small number is consistent with the standard model for colorectal tumorigenesis based on the identification of mutations in well-known cancer genes [6]. However, Sjöblom et al. [7] have recently determined the sequence of 13,000 genes in colorectal cancers and found that individual tumors contained an average of 62 nonsynonymous mutations. Extrapolating to the entire genome, it was estimated that individual colorectal cancers contain about 100 nonsynonymous mutations and that as many as 20 of the mutated genes in individual cancers might play a causal role in the neoplastic process [7].

Tumors arise from a process of replication, mutation, and selection through which a single cell acquires driver mutations which provide a fitness advantage by virtue of enhanced replication or resistance to apoptosis [8]. Each driver mutation thereby allows the mutant cell to go through a wave of clonal expansion. Along with drivers, passenger mutations, which do not confer any fitness advantage, are frequently observed. Passenger mutations arise in advantageous clones and become frequent by hitchhiking. The accumulation of ~100 mutations per cell is therefore the result of sequential waves of clonal expansion; the observed mutations mark the history of the cancer cell, including both drivers and passengers.

2 Genetic mutations can arise either due to errors during DNA replication or from exposure to

3 genotoxic agents. The normal mutation rate due to replication errors is in the range of $10^{-10}$ to

4 $10^{-9}$ per nucleotide per cell per division [9]. It is likely that the initial steps leading to cancer

5 arise in cells with a normal mutation rate [10]. A normal mutation rate might also be sufficient

6 to generate the large numbers of mutations in cancer given the many generations that the

7 dominant cancer cell clone has gone through both before and after its initiating mutation [11,12].

8 However, it has also been argued that tumor cells have mutator phenotypes that accelerate the

9 acquisition of mutations [13].

11 Mathematical modeling of carcinogenesis has had a rich history since its introduction more than

12 50 years ago [14,15,16]. The initial two-hit theory has evolved into more elaborate models

13 incorporating multiple hits, rate-limiting events and genomic instability [4,17,18,19,20,21,22].

14 Most models consider the stem cell at the base of the colonic crypt as the initial target for

15 mutation, with the daughter cells giving rise to the adenoma and progressively increasing the risk

16 of malignant development [4,21].

18 The tumor data collected by Sjöblom et al. [7] show that the mutational patterns among

19 colorectal cancers from different patients are diverse. This observation indicates that there may

20 be many different mutational pathways that can lead to the same cancer phenotype. In the model

21 described below, we assume that there are 100 potential driver genes and ask for the expected

22 waiting time until one cell has acquired mutations in a given number, up to 20, of these genes.

23 We assume that one or two initial mutations, perhaps together with losses or gains of large chromosomal regions [14,15], give rise to a benign tumor (adenoma) of ~1 milligram or $10^6$ cells (Fig. 1). We model the progression of this adenoma to full blown cancer over a period of 5 to 20 years [15], in which the adenoma grows to ~1 gram, or $10^9$ cells. Whether the whole population of cells is at risk for clonal expansion or whether a fraction of cells akin to stem cells drives growth of the adenoma is currently a subject of debate. This is important as cancer stem cells, as well as other factors such as geometric constraints on the architecture of the adenoma, may significantly reduce the effective population size and thereby impact the waiting time to cancer [23,24]. Note that it is not size that distinguishes a cancer from an adenoma; rather it is the ability of the cancer cells to invade through the underlying basement membrane and escape from its normal anatomical position.

We use the Wright-Fisher process [25] to model the somatic evolution of cancer in a colonic adenoma. We assume a cell turnover of one per day [26] and analyze the time to cancer as a function of the population size $N$, the per-gene mutation rate $u$, and the average selective advantage $s$ per mutation. We present extensive simulation results as well as analytical approximations to the expected waiting time. The model offers a basic understanding of how the different evolutionary forces contribute to the progression of cancer.

**Methods**

**Data.** The collection of tumor data has been described in [7]. Briefly, ~13,000 genes were sequenced from cancers of 11 patients with advanced colorectal cancers. Any mutant gene detected in this study was analyzed in an additional 24 patients with advanced cancers. Tumors with mismatch repair (MMR) deficiency were not included in this cohort, as MMR is known to increase the mutation rate by orders of magnitude and would complicate the analysis of mutations. Mutations were found in 519 genes and of these, 105 genes were found to be mutated in at least two independent tumors.

**Statistical analysis.** In order to test for dependencies between mutated genes, we calculated all 3003 pair-wise partial correlations between the 78 genes that were considered candidate drivers. Because the number of observed tumors is much smaller than the number of genes, we used the shrinkage method introduced in [27] for estimation.

**Wright-Fisher process.** We initially consider a colonic adenoma composed of $10^6$ cells (~1 mm$^3$) that is growing exponentially to reach a size of $10^9$ cells (~1 cm$^3$). Serial radiological observations show that the growth of unresected colonic adenomas is well approximated by an exponential function [28]. The average growth rate determined in [28] implies that it takes ~11 years for an adenoma to grow from $10^6$ to $10^9$ cells. We consider an evolving cell population of size $N(t)$ in generation $t$. Population growth is modeled by assuming that growth is proportional to the average fitness ‹w› of the population, $N(t+1) - N(t) = \alpha$ ‹w› $N(t)$, where $\alpha$ is a constant ensuring the experimentally observed growth dynamics, and $N(0) = 10^6$. Although ‹w› changes slightly over time, the growth kinetics is still approximately exponential.

Each cell is represented by its genotype, which is a binary string of length $d = 100$ corresponding to the 100 potential driver genes. The population is initially homogeneous and composed of wild type cells which are represented by the all-zeros string. In each generation, $N(t)$ genotypes are sampled with replacement from the previous generation. For large population sizes of $10^9$ cells, it 1  is not feasible to track the fate of each of the possible $2^{100}$ mutants in computer simulations.

2  However, we are interested in the first appearance of any *k*-fold mutant in the system ($k = 20$).

3  Thus, it suffices to trace the $k + 1$ mutant error classes, i.e., the number of *j*-fold mutants $N_j(t)$ for

4  each $j = 0, \ldots, k$, in each generation. With every additional mutation we associate a selective

5  advantage *s*. Thus, the relative fitness of a *j*-fold mutant is $w_j = (1+s)^j \big/ \sum_{i=0}^{k}(1+s)^i x_i$, where

6  $x_i = N_i/N$, and the average population fitness is ‹w› = $\sum_{j=0}^{k} x_j w_j$. Ignoring back mutation, the

7  probability of sampling a *j*-fold mutant is

8  $$\theta_j = \sum_{i=0}^{j} \binom{d-i}{j-i} u^{j-i}(1-u)^{d-j} w_j x_i(t),$$

9  where *u* is the mutation rate per gene. In each generation, the population is updated by sampling

10 from the multinomial distribution

11 $$(N_0(t+1),\ldots,N_k(t+1)) \sim \frac{N(t)!}{N_0(t)! \cdot \ldots \cdot N_k(t)!} \prod_{j=0}^{k} \theta_j^{N_j(t)},$$

12 where $N(t)$ follows the above growth kinetics.

14 We use the discrete Wright-Fisher process rather than the continuous Moran process [25], which

15 might seem more natural for cancer progression, because the Wright-Fisher process allows for

16 efficient computer simulations even for very large population sizes. Both models behave

17 similarly for large population sizes [25].

19 **Analytical approximation.** The large cell population size might suggest that one could consider

20 a replicator equation in the limit as $N \to \infty$. However, this approach yields a Poisson distribution

Beerenwinkel *et al.*for the time-dependent relative frequencies $x_j(t)$ with parameter $\lambda = ud(e^{st}-1)/s$, implying that the variance of $x$ increases over time, which contrasts with the simulation results (Fig. 3). The reason for this discrepancy is that, in the replicator equation, higher order mutants with high fitness are instantaneously generated. Thus, the time for their expansion is underestimated compared to the waiting time in the stochastic system. See supporting information for further discussion of this phenomenon.

In order to account for the stochastic fluctuations in the accumulation of $k$ mutations, we model this process by decoupling mutation and selection (see supporting information S3 for mathematical details). Briefly, we assume that $j$-fold mutants are generated at a constant rate with increasing $j$. The Gaussian describing the distribution of mutant error classes has mean $vt$, variance $\sigma^2$, and travels with velocity $v = s\sigma^2$ (Fig. 3). In order to determine $v$, we consider an (initially) exponentially growing subpopulation of $j$-fold mutants and calculate the expected time until one $(j+1)$ mutant is produced. This leads to $v = 2s\log N \Big/ \left(\log \dfrac{s}{ud}\right)^2$ and for constant population size $N$, we obtain the approximation $t_k \approx k\left(\log \dfrac{s}{ud}\right)^2 \Big/ 2s\log N$ for the expected time to the first appearance of any $k$-fold mutant. The same waiting time in a population growing exponentially from initial size $N_{\text{init}} = N(0)$ to final size $N_{\text{fin}} = N(t_k)$ is equal to that in a constant population with effective population size $N = \sqrt{N_{\text{init}} N_{\text{fin}}}$. Thus the speed of the mutant wave in the growing population can be approximated by the average of the values corresponding to the initial and final population sizes. This leads to $t_k \approx k\left(\log \dfrac{s}{ud}\right)^2 \Big/ s\log(N_{\text{init}} N_{\text{fin}})$ for the waiting time in a population growing from $N_{init}$ to $N_{fin}$. We will often restrict our attention to constant population sizes because of the equivalent waiting time in a constant population with effective size equal to the geometric mean of the initial and final population sizes.

**Results**

The mutation data are represented in a binary matrix of size $35 \times 78$ whose rows correspond to 35 tumor samples and whose columns correspond to the 78 candidate cancer genes identified by Sjöblom et al. [7] (Fig. 2). A non-zero entry in cell $(i, j)$ of this matrix indicates the presence of a mutation in gene $j$ of tumor $i$. Tumors harbor between 1 and 20 mutated genes (mean = 6.5). Most of these genes (66/78 = 85%) are mutated in at most 3 different tumors resulting in highly diverse mutational patterns among the tumors. The notable exception are the three well-known cancer genes *APC*, *p53*, and *K-ras* which were found mutated in 24, 17, and 16 tumors, respectively. We have analyzed partial correlations between genes taking into account the small number of observations and multiple comparisons. Several pairs of genes were significantly correlated, most of them positively, but all correlations were weak and below 0.07 (Fig. S1). From this data analysis, we conclude that in colon cancer, a very small number of genes are mutated frequently and are present in many tumors. However, many other genes are involved in tumor progression, although each single gene is mutated only in a small subset of tumors without a clear pattern emerging.

For the purpose of mathematical modeling of tumorigenesis, we consider the presence of an adenoma. Adenoma formation probably requires the appearance of mutations in a few genes (e.g., *APC* and *K-ras*) that are common to most tumors. We assume the occurrence of all

Beerenwinkel *et al.*Beerenwinkel *et al.*

1  subsequent mutations to be independent events. When any $k$ out of $d = 100$ susceptible genes are
2  mutated in a single cell, the cancer phenotype is considered to be attained. The first cells of this
3  type mark the onset of an invasive tumor. The Wright-Fisher process is used to describe these
4  evolutionary dynamics. Despite the large population size of up to $N = 10^9$ cells, we can
5  efficiently compute estimates of the time to the first appearance of any $k$-fold mutant by
6  simulation, because it suffices to trace the distribution of the $k + 1$ mutant error classes in each
7  generation. We assume a constant average selective advantage, $s$, for each mutation and a per-
8  gene mutation rate, $u$. Figure 3 displays the typical behavior of this process in a single
9  simulation. After a short initial phase in which the homogeneous wild type population produces
10 the first low-order mutants, a traveling wave is observed (Fig. 3). Apparently, this distribution of
11 error classes has constant variance and travels with constant velocity towards higher-order
12 mutants. Thus, we expect the time until the first $k$-fold mutant appears to be linear in $k$. This
13 conjecture is substantiated by simulations for a wide range of parameters (Fig. S2) provided that
14 mutations are advantageous ($s > 0$).

16 Within our model, the probability of developing cancer is equated with the probability of
17 generating at least one $k$-fold mutant cell in the adenoma. For $k = 20$, this probability as a
18 function of time is depicted in Figure 4. The expected time to the development of cancer
19 increases with decreasing cell population size (hence the low risk of cancer associated with very
20 small adenomas), with decreasing selective advantage, and with decreasing mutation rate. Thus,
21 if the population at risk is a small subset composed of actively replicating stem cells, tumor
22 progression will be slow.  In contrast, an increased mutation rate due to genetic instability speeds
23 up this process.

Page 10 of 30

2  The simulations suggest that in a time frame of 5 to 15 years, cancer might develop in an
3  adenoma of size $10^7$ to $10^9$ cells with a normal mutation rate of $10^{-7}$ per gene per cell division
4  and a 1% selective advantage per mutation (Fig. 4a). Alternatively, a higher mutation rate of $10^{-5}$
5  per gene per cell division would enable a smaller population of at-risk cells ($10^5$ to $10^7$) and a
6  smaller selective advantage (0.1%) to reach the required number of mutations in the same time
7  interval (Fig. 4b). However, for reasonable mutation rates, a completely neutral process ($s = 0$)
8  predicts waiting times that are not consistent with the observed incidence of colon cancer, as
9  would be expected (Fig, 4, Fig. S2).

11 Figure 5 generalizes these findings to different values of *k* by partitioning the parameter space of
12 the model into regions of identical evolutionary outcomes. Each curve defines an instance of the
13 Wright-Fisher process that results in a 10% chance of developing a *k*-fold mutant after 3000
14 generations (or 8.2 years). These level curves define the parameter combinations that produce
15 similar dynamics. For example, a small at-risk population is unlikely to generate a cancer
16 requiring more than 10 driver gene mutations unless the selective advantage for these mutations
17 is large (see Discussion).

19 Based on the simulation results we have derived an analytical approximation for the expected
20 time to cancer. The key observation is that the distribution of error types follows a Gaussian (Fig.
21 3). This approach leads to the expression

22 (1) $$t_k = k \frac{\left(\log \frac{s}{ud}\right)^2}{s \log(N_{init} N_{fin})}$$

1  for the expected waiting time, where $k$ is the number of cancer-defining genes, $d$ is the number of
2  susceptible genes, $u$ is the mutation rate, $s$ the average selective advantage, and $N_{init}$ and $N_{fin}$
3  are the initial and final population sizes of the polyp, respectively (see Materials and Methods).
4  The approximation is linear in $k$ (Fig. S2) and matches closely the observed behavior of the
5  Wright-Fisher process, as long as $s > 0$ (Fig. 3-5). The fit is analyzed quantitatively in the
6  supporting information S3. The expression for $t_k$ highlights the strong effect of the selective
7  advantage on tumorigenesis, and gives an explicit tradeoff between the evolutionary forces.
8
9  **Discussion**
10 Research over the past three decades has shown that cancer is an acquired genetic disorder [1].
11 The process of replication, mutation, and selection eventually leads to the appearance of tumors
12 in multi-cellular organisms if they live long enough. Tumor cells accumulate many mutations in
13 their evolutionary path [7,8,29], but not all mutations play a causal role in the evolution of the
14 clone.  If a gene is mutated in tumors derived from different patients, it is less likely to be a
15 passenger and more likely to provide the cell with a selective advantage to expand and dominate
16 the population. Based on this reasoning, Sjöblom et al [7] suggested that as many as ~20 driver
17 genes are mutated per tumor. The diverse mutational landscapes observed in tumor cells of the
18 same tissue origin suggest that different mutations can have the same phenotypic effect. One
19 plausible explanation for this observation is that genes are organized into intracellular pathways
20 (signaling, metabolic, checkpoint etc) and the disturbance of these pathways drives
21 tumorigenesis. Within each cell, every information transfer cascade requires functional proteins
22 that are the products of distinct genes. Mutations in any one of the genes that code for proteins in
23 a given pathway can complement each other and their genetic alterations can have similar phenotypic effects [1]. This view is supported by the observation that multiple hits in different genes of the same pathway in individual tumors are less frequent than expected [1].

In our model, we assume that each subsequent mutation has the same incremental effect on the fitness of the cell. In general, however, the impact of a specific mutation on the phenotype of the cell will depend on the genetic background. Gene interactions, or epistasis, can be positive or negative, and they can impose constraints on the order in which mutations accumulate [1]. In this case, the model parameter $s$ may be regarded as the average fitness increase per mutation. In another simplifying abstraction, we have defined the tumor cell by the accumulation of $k = 20$ mutations in different driver genes. In reality, it is unlikely that any combination of 20 genes will induce the cancer phenotype. Our assumption is based on the observed cancer genotypes which fail to reveal a striking genetic signature of cancer cells. In this respect, our model provides lower bounds on the expected waiting time to cancer, as reaching a specific 20-fold mutant may take significantly longer.

These abstractions are important because all lesions begin with a small number of neoplastic cells. The simulations in Figure 5 show that cancers would never result from such small numbers of cells if 20 driver mutations were required and each mutation conferred only a small fitness advantage. It is likely that some of the early mutations increase fitness more than the average, allowing a small, initiating lesion to grow into an intermediate size lesion. Once a growth reaches this size, mutations with small fitness advantages can accumulate and eventually convert the tumor into a cancer.

1  The large population size of $10^9$ cells would suggest that a purely deterministic approximation to
2  the Wright-Fisher process is reasonable. It turns out, however, that the stochasticity associated
3  with generating mutants of each new type has a strong impact on the evolutionary dynamics (see
4  supporting information S3). Therefore, a deterministic model of evolutionary dynamics will
5  significantly underestimate the time to cancer. The closer approximation presented here exploits
6  the regular behavior of the system of propagating a Gaussian distribution of error types and takes
7  into account stochastic effects in determining the speed of this traveling wave. Thus, stochastic
8  effects can play an important role even in very large populations.
9
10 Tumors derived from the same tissue exhibit considerable variability in their spectrum of
11 mutations (Fig. 2 and [7]). The number and type of mutations observed is the result of the size of
12 the population at risk, the mutation rate, and the microenvironment of the evolving clone. The
13 individual mutation rate can vary significantly due to genetic [29,30], and environmental effects
14 (e.g., dietary fat intake, colonic bacterial flora, prior genotoxic therapy) [31,32]. These factors
15 that are expected to be different for every tumor also contribute to the diversity of the mutational
16 landscape observed in tumors. It is also worth noting that the number of potential driver genes is
17 likely to be an underestimate because the power of the Sjöblom et al. study to detect infrequent
18 mutations was limited [7]. The study of larger numbers of tumors is likely to show that a few
19 hundred different genes may function as drivers. This increase in potential drivers, however,
20 will not have a substantial effect on the conclusions of the models derived here (Eq. 1).
21
22 Most tissues in metazoans undergo turnover and are maintained by a population of tissue specific
23 stem cells that generally replicate at a slow rate and exhibit properties such as asymmetric division and immortal DNA strand co-segregation [33], perhaps to minimize the acquisition and retention of mutations. Although many tumors have cancer stem cells at their root [34] and colon cancer stem cells have been reported [23,24], it is an open question whether such cells arise solely due to the progressive accumulation of mutations in normal stem cells, or because cells can re-acquire stem cell-like properties by mutation. The former scenario would suggest a much smaller effective population size, an important variable for modeling the evolution of cancer [4,21,35,36,37]. The colon has approximately $10^7$ crypts, each one maintained by a small number of stem cells [26]. Initially, these stem cells constitute the overall population at risk, but the vast majority of patients with colon cancer develop tumors as the natural progression of mucosal adenomas [38]. Thus, adenoma formation can be regarded as a mechanism by which the population of cells at risk is increased and hence the probability of cancer in patients with multiple adenomas is dramatically increased. This is observed in familial adenomatous polyposis patients, who have inherited mutations of the *APC* gene.

Our model permits investigation of the impact of the relevant parameters of tumor evolution on a global scale. These parameters include the size of the population at risk, the mutation rate, and the fitness advantage conferred by specific mutations (Eq. 1). The model suggests that the average waiting time for the appearance of the tumor is strongly affected by the fitness, *s*, conferred by the mutations, with the average waiting time decreasing roughly as 1/*s* (Fig. S2). The mutation rate and the size of the population at risk contribute only logarithmically to the waiting time and hence have a weaker impact. Thus, the model of cancer progression presented here might add to the debate whether selection [10,11] or mutation [39] is the dominant force in tumor development.

Finally, this model helps answer several questions about colorectal tumorigenesis that have long perplexed researchers and clinicians. Why is there so much heterogeneity in the times required for tumor progression among different patients? Why is there so much heterogeneity in the sizes and development times of tumors even within individual patients, such as those with familial adenomatous polyposis, if they all have the same initiating *APC* mutation? Why do cancers behave so differently with respect to their response to chemotherapeutic agents or radiation or their propensity to metastasize? Our model is compatible with the view that a few major mutational pathways, such as those involving *APC*, *K-ras*, and *p53*, endow relatively large increases in fitness that can allow tumors to grow to sizes compatible with further progression (Fig. 5). However, the final course to malignancy will be determined by multiple mutations, each with a small and distinct fitness advantage, and these mutations occur stochastically. Every cancer will thereby be dependent on a unique complement of mutations that will determine its propensity to invade, its ability to metastasize, and its resistance to therapies. If this model is correct, then biological heterogeneity is a direct consequence of the tumorigenic process itself.

Beerenwinkel *et al.*26. Potten CS (1998) Stem cells in gastrointestinal epithelium: numbers, characteristics and death. Philos Trans R Soc Lond B Biol Sci 353: 821-830.
27. Schafer J, Strimmer K (2005) A shrinkage approach to large-scale covariance matrix estimation and implications for functional genomics. Stat Appl Genet Mol Biol 4: Article32.
28. Welin S, Youker J, Spratt JS, Jr. (1963) The Rates and Patterns of Growth of 375 Tumors of the Large Intestine and Rectum Observed Serially by Double Contrast Enema Study (Malmoe Technique). Am J Roentgenol Radium Ther Nucl Med 90: 673-687.
29. Stoler DL, Chen N, Basik M, Kahlenberg MS, Rodriguez-Bigas MA, et al. (1999) The onset and extent of genomic instability in sporadic colorectal tumor progression. Proc Natl Acad Sci U S A 96: 15121-15126.
30. Friedberg EC (2003) DNA damage and repair. Nature 421: 436-440.
31. Slattery ML, Curtin K, Anderson K, Ma KN, Edwards S, et al. (2000) Associations between dietary intake and Ki-ras mutations in colon tumors: a population-based study. Cancer Res 60: 6935-6941.
32. Brink M, Weijenberg MP, De Goeij AF, Schouten LJ, Koedijk FD, et al. (2004) Fat and K-ras mutations in sporadic colorectal cancer in The Netherlands Cohort Study. Carcinogenesis 25: 1619-1628.
33. Rambhatla L, Ram-Mohan S, Cheng JJ, Sherley JL (2005) Immortal DNA strand cosegregation requires p53/IMPDH-dependent asymmetric self-renewal associated with adult stem cells. Cancer Res 65: 3155-3161.
34. Reya T, Morrison SJ, Clarke MF, Weissman IL (2001) Stem cells, cancer, and cancer stem cells. Nature 414: 105-111.
35. van Leeuwen IM, Byrne HM, Jensen OE, King JR (2006) Crypt dynamics and colorectal cancer: advances in mathematical modelling. Cell Prolif 39: 157-181.
36. Michor F, Hughes TP, Iwasa Y, Branford S, Shah NP, et al. (2005) Dynamics of chronic myeloid leukaemia. Nature 435: 1267-1270.
37. d'Onofrio A, Tomlinson IP (2007) A nonlinear mathematical model of cell turnover, differentiation and tumorigenesis in the intestinal crypt. J Theor Biol 244: 367-374.
38. Winawer SJ (1999) Natural history of colorectal cancer. Am J Med 106: 3S-6S; discussion 50S-51S.
39. Bielas JH, Loeb KR, Rubin BP, True LD, Loeb LA (2006) Human cancers express a mutator phenotype. Proc Natl Acad Sci U S A 103: 18238-18242.Page 18 of 30

**Acknowledgments**

We are grateful to Tobias Sjöblom, Sian Jones, Jimmy Lin, Laura Wood, and Yoh Iwasa for helpful discussions.

**Funding**

Support from the NSF/NIH joint program in mathematical biology (NIH grant GM078986) is gratefully acknowledged. Genomics studies at Johns Hopkins are funded by the Virginia and D.K. Ludwig Fund for Cancer Research, the National Colorectal Cancer Research Alliance, The Maryland Tobacco Fund and NIH grants CA43460, CA57345, CA62924, and CA121113. N. B. is funded by a grant from the Bill & Melinda Gates Foundation through the Grand Challenges in Global Health Initiative. The Program for Evolutionary Dynamics at Harvard University is sponsored by Jeffrey Epstein.

**Author Contributions**

K.W.K. V.E.V., and B.V. performed experiments; N.B., T.A., D.D., A.T., and M.A.N. developed the mathematical model; and N.B., T.A., D.D., B.V., and M.A.N. wrote the paper.

**Competing Interests**

We declare that no competing interests exist.

**Abbreviations**

*APC*, adenomatous polyposis coli gene; *K-ras*, Kirsten rat sarcoma 2 viral oncogene homolog; MMR, mismatch repair; *p53*, tumor protein 53

1   **Figures**

3   Figure 1

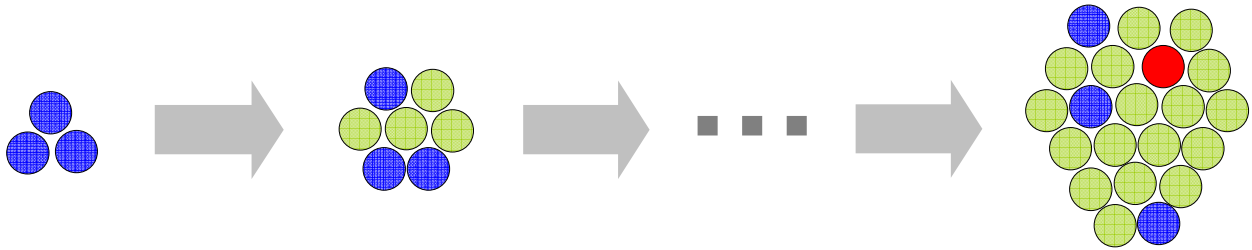

1    Figure 2

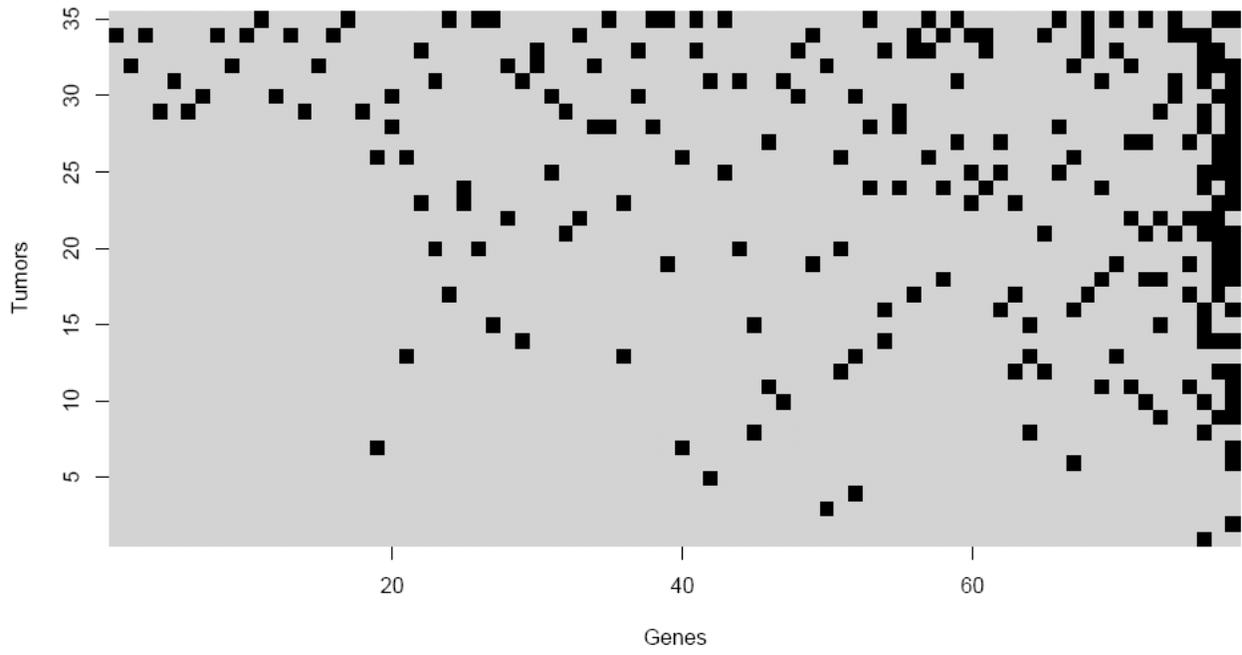

1   Figure 3

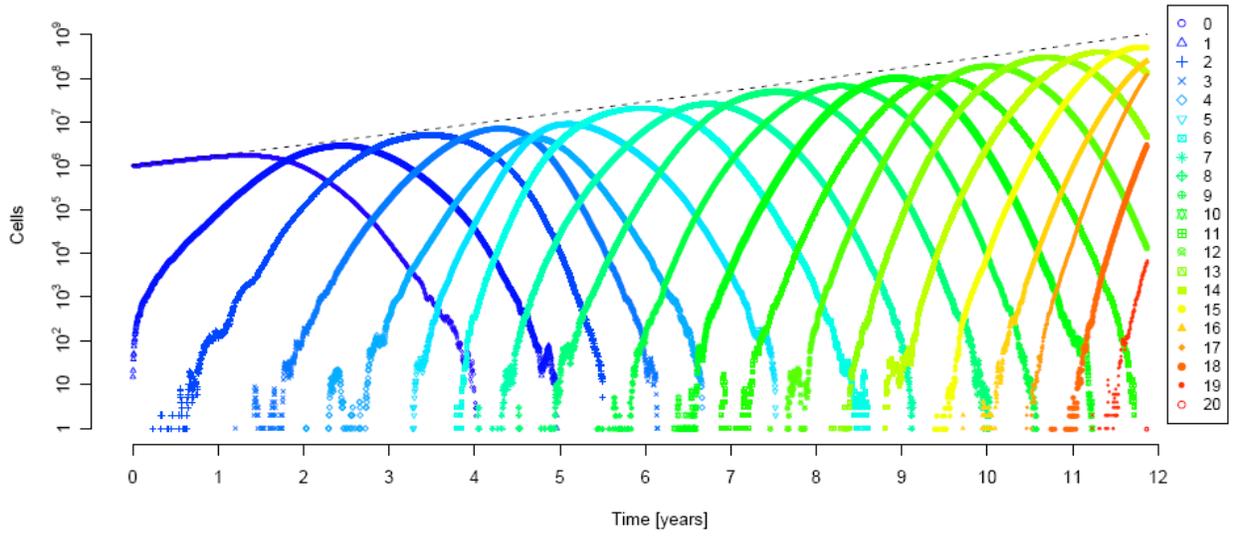

Figure 4

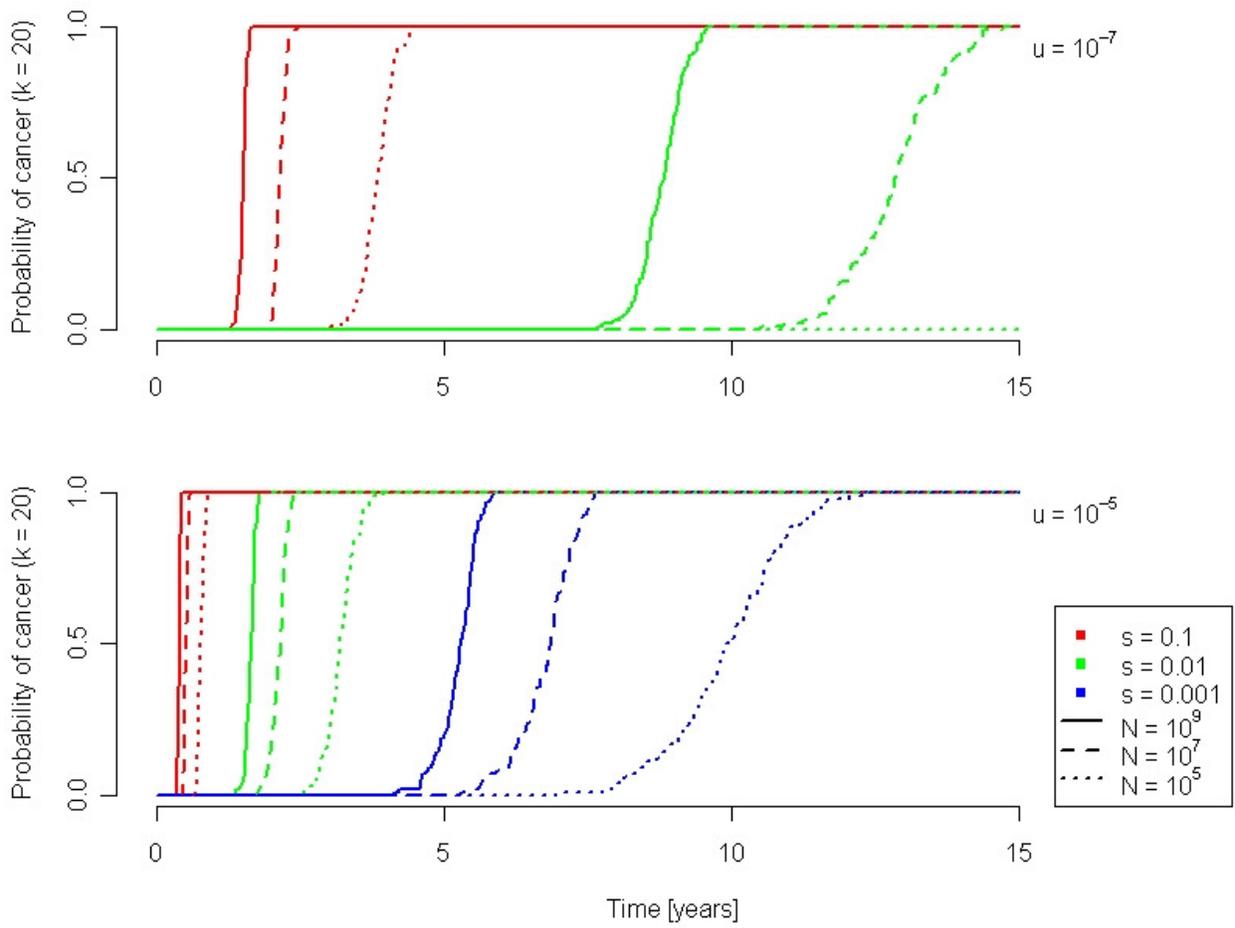

1  Figure 5

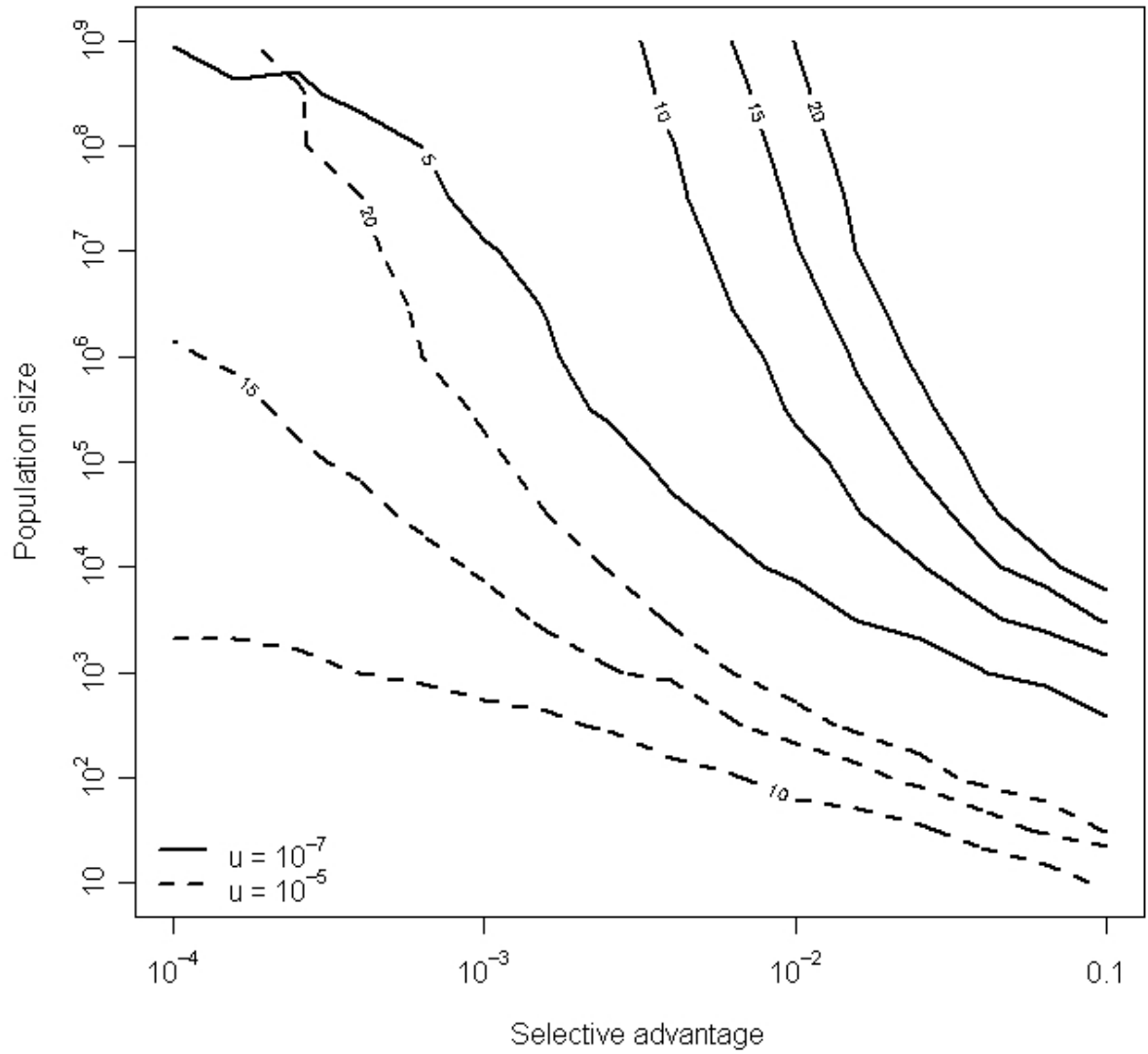

**Figure Legends**

**Figure 1.** Schematic representation of the evolution of cancer in a colonic adenoma. The adenoma grows from a population of $10^6$ to $10^9$ cells which accumulate mutations that drive phenotypic changes seen in cancer cells. Blue circles symbolize adenoma cells prior to accumulating the additional mutations that are the subject of modeling, green indicates cells that have acquired additional, but an insufficient number of mutations for malignancy, and red indicates cells with the number of mutations required for the cancer phenotype.

**Figure 2.** Mutational patterns in 35 late-stage colorectal cancer tumors from Sjöblom et al. (2006). Matrix rows are indexed by tumors, columns are indexed by cancer-associated genes as identified by Sjöblom et al. (2006). Dark spots indicate mutated genes. Both tumors and genes have been sorted by an increasing number of mutations. The three genes mutated most often are *APC* (in 24 tumors; last column), *p53* (in 17 tumors; penultimate column), and *K-ras* (in 16 tumors; adjacent to *p53* column).

**Figure 3.** Evolution of cancer modeled by the Wright-Fisher process. The distribution of cells in the error classes $N_0, \ldots, N_{20}$ is displayed in a single simulation over a time period of 12 years after which the first cell harboring 20 mutations appears. The total population size (dashed line) grows exponentially from $10^6$ to $10^9$ cells in this time period, Each cell has 100 susceptible genes, all of which are of wild type initially. We further assumed a mutation rate of $10^{-7}$ per gene, a 1% selective advantage per mutation, and a turnover of 1 cell division per cell per day.

Beerenwinkel *et al.*Each error class has an approximately Gaussian distribution (after a short initial phase), but the introduction of each new mutant is subject to stochastic fluctuations.

**Figure 4.** The probability of developing cancer, defined as the occurrence of a cell with any 20 mutated genes out of 100. Simulation results are displayed for three different population sizes ($10^9$, solid lines; $10^7$, dashed lines; $10^5$, dotted lines), three different selection coefficients (10%, red lines; 1%, green lines; 0.1%, blue lines), and two different mutation rates ($10^{-7}$, top; $10^{-5}$, bottom).

**Figure 5.** Level curves of identical cancer dynamics. Each curve connects points in parameter space (*x*-axis: selective advantage *s*, *y*-axis: population size *N*) with the same evolutionary outcome, namely a 10% chance of developing a *k*-fold mutant after 8.2 years (or 3000 generations). The mutation rate is $10^{-7}$ (solid lines) and $10^{-5}$ (dashed lines), respectively. Curves are labeled with the number *k* of mutated genes that defines the cancer phenotype.

1   **Supporting information**

3   **Figure S1.** Histogram of $3003 = \binom{78}{2}$ partial correlations between all 78 cancer-associated

4   genes. Correlation coefficients have been computed from the 0/1 matrix displayed in Figure 2.

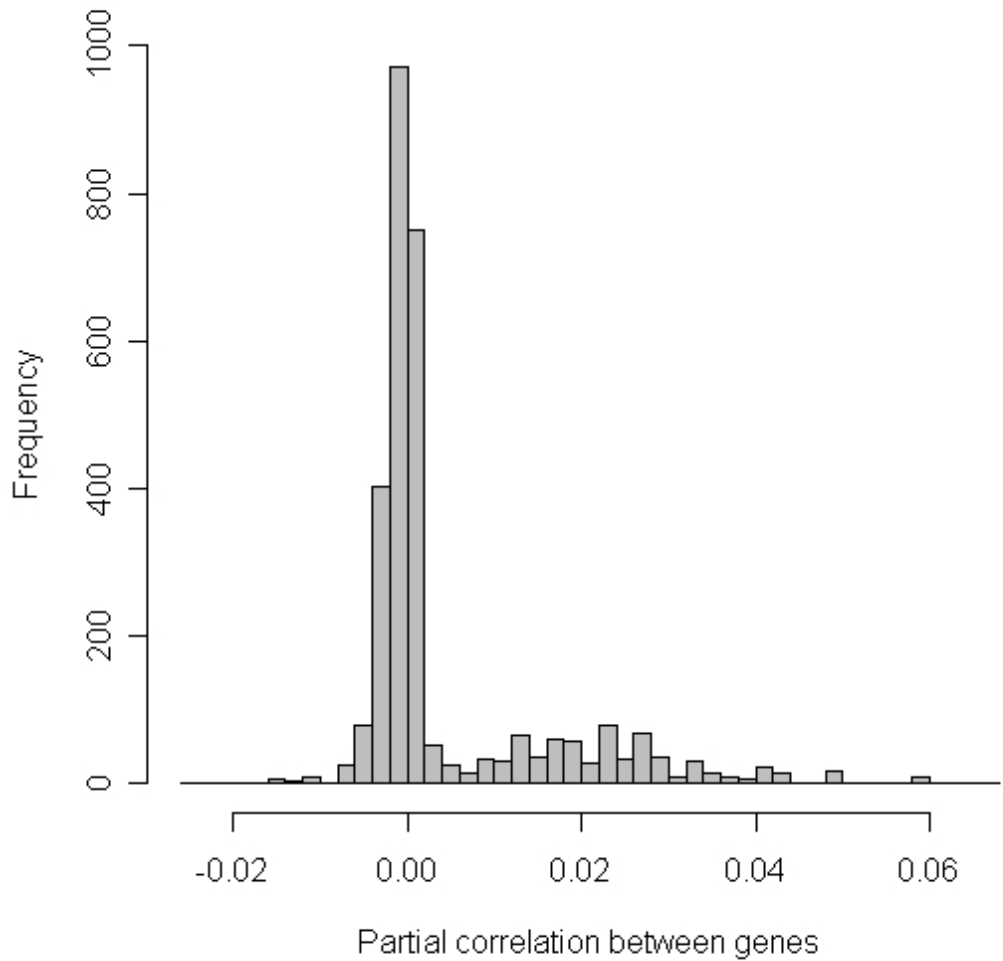

1   **Figure S2.** Time $T_k$ until, in 10% of patients, $k$ genes are mutated. The waiting time $T_k$ (*y*-axis) is
2   plotted versus the number $k$ of mutated genes (*x*-axis). Left panels correspond to a normal
3   mutation rate of $u = 10^{-7}$, right panels to an increased mutation rate of $u = 10^{-5}$. Population sizes
4   of $10^5$ (top panels), $10^7$ (middle panels), $10^9$ (bottom panels) are considered. The selective
5   advantage per mutation varies among 0.1 (red lines), 0.01 (green), 0.001 (cyan), and 0 (blue).

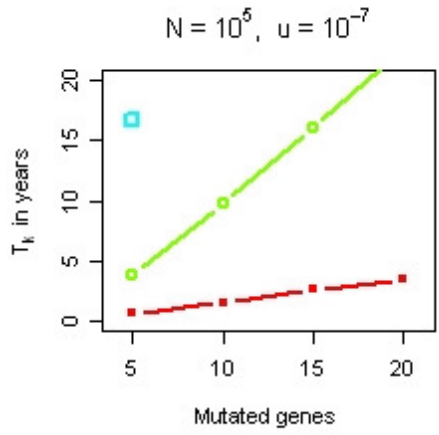
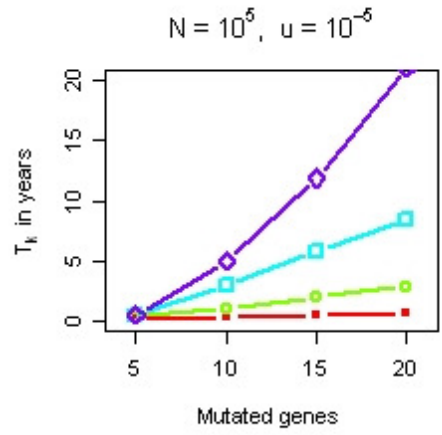
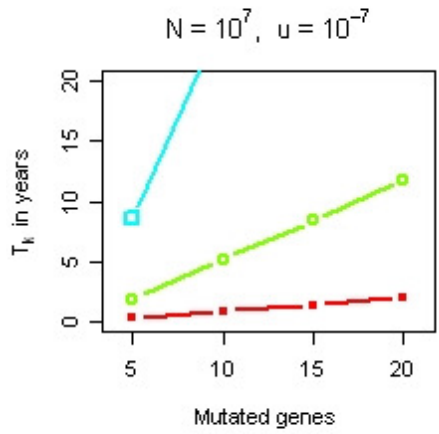
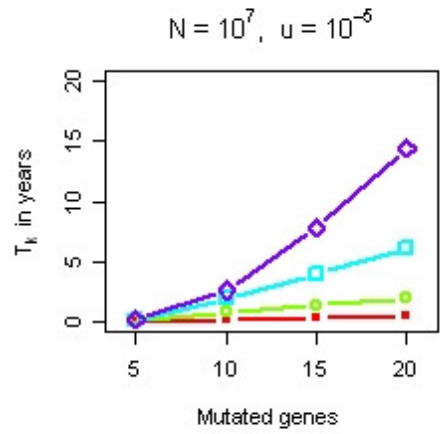
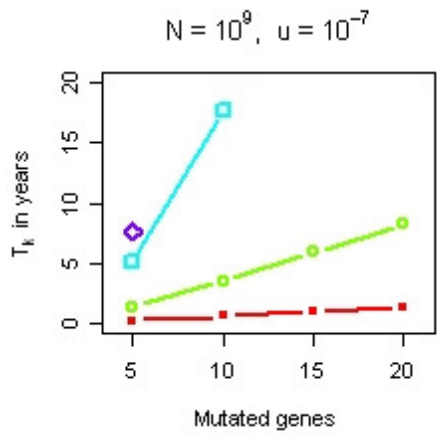
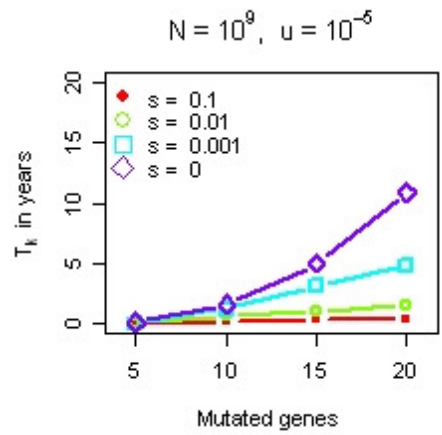

**Supporting document S3**

PDF document entitled "Analytical approximation for the expected waiting time" which contains the mathematical details of the model.